\let\TeXyear\year
\documentclass{ieeeaccess}

\let\year\TeXyear
\usepackage{cite}
\usepackage{amsmath,amssymb,amsfonts}
\usepackage{algorithmic}
\usepackage{graphicx}
\usepackage{textcomp}

\usepackage{bm}
\makeatletter
\AtBeginDocument{\DeclareMathVersion{bold}
\SetSymbolFont{operators}{bold}{T1}{times}{b}{n}
\SetSymbolFont{NewLetters}{bold}{T1}{times}{b}{it}
\SetMathAlphabet{\mathrm}{bold}{T1}{times}{b}{n}
\SetMathAlphabet{\mathit}{bold}{T1}{times}{b}{it}
\SetMathAlphabet{\mathbf}{bold}{T1}{times}{b}{n}
\SetMathAlphabet{\mathtt}{bold}{OT1}{pcr}{b}{n}
\SetSymbolFont{symbols}{bold}{OMS}{cmsy}{b}{n}
\renewcommand\boldmath{\@nomath\boldmath\mathversion{bold}}}
\makeatother

\def\BibTeX{{\rm B\kern-.05em{\sc i\kern-.025em b}\kern-.08em
    T\kern-.1667em\lower.7ex\hbox{E}\kern-.125emX}}

\usepackage{threeparttable}
\usepackage{caption}
\usepackage{soul,color,ulem} 
\usepackage{microtype}
\usepackage{booktabs}
\usepackage{float}

\usepackage[commandnameprefix=always,final]{changes} 

\definecolor{accessblue}{cmyk}{1,0.3,0,0.2}
\definecolor{greycolor}{cmyk}{0,0,0,.8}

\begin{document}

\history{Date of publication xxxx 00, 0000, date of current version xxxx 00, 0000.}
\doi{10.1109/ACCESS.2026.xxxxxxx}

\title{Parameter-Efficient Transfer Learning for Microseismic Phase Picking Using a Neural Operator}
\author{\uppercase{Ayrat Abdullin}\authorrefmark{1},
\uppercase{Umair Bin Waheed}\authorrefmark{1}, \uppercase{Leo Eisner}\authorrefmark{2}, AND \uppercase{Naveed Iqbal}\authorrefmark{3}}

\address[1]{Department of Geosciences, King Fahd University of Petroleum and Minerals, Dhahran, 31261, Saudi Arabia}
\address[2]{Seismik s.r.o., Prague, 18200, Czech Republic}
\address[3]{Department of Electrical Engineering, King Fahd University of Petroleum and Minerals, Dhahran, 31261, Saudi Arabia}
\tfootnote{This work was supported in part by the Deanship of Research (DR) at King Fahd University of Petroleum \& Minerals (KFUPM) through project No. MbSC2601.}

\corresp{Corresponding author: Ayrat Abdullin (e-mail: g202203180@kfupm.edu.sa).}

\begin{abstract}
Seismic phase picking is fundamental for microseismic monitoring and subsurface imaging. Manual processing is impractical for real-time applications and large sensor arrays, motivating the use of deep learning-based pickers trained on extensive earthquake catalogs. On a broader scale, these models are tuned to perform optimally in high signal-to-noise and long-duration networks and fail to perform satisfactorily when applied to campaign-based microseismic datasets, which are characterized by low signal-to-noise ratios, sparse geometries, and limited labeled data.

In this study, we present a microseismic adaptation of a network-wide earthquake phase picker, Phase Neural Operator (PhaseNO), using transfer learning and parameter-efficient fine-tuning. Starting from a model pre-trained on more than 57,000 three-component earthquake and noise records, we fine-tune it using only 200 labeled and noisy microseismic recordings from hydraulic fracturing settings. We present a parameter-efficient adaptation of PhaseNO that fine-tunes a small fraction of its parameters (only 3.6\%) while retaining its global spatiotemporal representations learned from a large dataset of earthquake recordings.

We then evaluate our adapted model on three independent microseismic datasets and compare its performance against the original pre-trained PhaseNO, a STA/LTA-based workflow, and two state-of-the-art deep learning models, PhaseNet and EQTransformer. Our adapted model significantly outperforms the original PhaseNO in F1 and accuracy metrics, achieving up to 30\% absolute improvements in all test sets and consistently performing better than STA/LTA and state-of-the-art models. With our adaptation being based on a small calibration set, our proposed workflow is a practical and efficient tool to deploy network-wide models in data-limited microseismic applications.

\chadded{The code is provided at https://github.com/SEEM-KFUPM/MicroPhaseNO.}
\end{abstract}


\titlepgskip=-15pt

\maketitle


\section*{Introduction}

Earthquake detection and localization are critical for real-time monitoring of induced seismic activity. Single-station characteristic-function pickers can provide approximate arrival times. To achieve reliable event localization and interpretation, it is crucial to obtain consistent multi-station phase picks either manually or through automated picking and association. Traditionally, seismologists rely on visual inspection of waveforms for multiple stations to manually identify P- and S-wave arrival times. This is subjective and prone to errors. These limitations have motivated the development of alternatives in automated phase picking methods, which increasingly rely on deep learning techniques.

Recently, advancements in neural network-based approaches have significantly improved phase picking efficiency~\cite{ross2018generalized, zhu2019phasenet, mousavi2020earthquake, baker2021monitoring, feng2022edgephase, shi2024labquakes}. These techniques have often been used in their original form~\cite{van2021oksp, retailleau2022wrapper} or after fine-tuning with custom data sets~\cite{chai2020using, armstrong2023deep}. For instance, Ross et al. (2018) proposed a CNN approach for accurately determining phase arrival times and P-wave polarity with high precision compared to expert human seismologists~\cite{ross2018generalized}. They used a custom data set of Southern California Seismic Network data. Zhu and Beroza (2019) presented PhaseNet, which is a fully convolutional neural network for robust P- and S-wave arrival detections in complex environments~\cite{zhu2019phasenet}. Wang et al. (2019) achieved state-of-the-art performance with a CNN approach and used high sensitivity three-component data in Japan~\cite{wang2019deep}. Chai et al. (2020) used pre-trained networks for phase picking in laboratory hydraulic fracturing experiments~\cite{chai2020using}. Recently, Zhou et al. (2025) proposed AI-PAL, which is a Self-Attention Recurrent Neural Network combined with rule-based detections of phase picking, phase association, and event localization~\cite{zhou2025ai}.

These studies show that phase picking using deep learning techniques is effective in low signal-to-noise ratio conditions and has been successfully used in various experiments for monitoring earthquakes and induced microseismic events, often benefiting from data augmentation and transfer learning strategies~\cite{anikiev2023machine}. Prior studies have shown that models trained on large tectonic datasets can generalize to laboratory experiments~\cite{chai2020using}, hydraulic fracturing environments~\cite{zhang2022phase}, and sparse deployments after fine-tuning, even with limited labeled data. However, most existing approaches operate on a single-station basis, processing each waveform independently and relying on post hoc association across the network.

This single-station design also remains dominant in prior microseismic and mining studies, including those based on architectures other than CNNs. For example, sequence models such as LSTM were used for arrival picking by Kirschner et al. (2019)~\cite{kirschner2019detecting}, demonstrating the effectiveness of local earthquake classification. He et al. (2020) employed capsule networks for P-waves detection in copper mine monitoring, achieving better generalization with fewer training samples than CNNs~\cite{he2020pickcapsnet}. Johnson et al. (2021) adapted Ross et al.'s CNN algorithm~\cite{ross2018generalized} for microseismic surveying of mining, showing improved performance after fine-tuning on a specific dataset~\cite{johnson2021application}. Kolar et al. (2023) developed an RNN-based phase picker, adapted for data from a local seismic monitoring array designed for the analysis of induced seismicity~\cite{kolar2023arrival}. Most machine learning (ML) phase-picking algorithms are based on single-station strategies, which may result in missed detection of low SNR events or false detection of local noise signals with emergent arrivals. 

Recently, array-based arrival picking algorithms have been proposed that exploit the spatial seismic phase coherence across multiple stations. For example, Chen and Li (2022) demonstrated that the CubeNet model effectively suppresses local spurious noise that often compromises trace-based methods~\cite{chen2022cubenet}. Similarly, Feng et al. (2022) presented EdgePhase, a multi-station arrival picking algorithm integrating the Edge Convolutional module with EQTransformer~\cite{feng2022edgephase}. In comparison with the standard EQTransformer, EdgePhase demonstrated a 5\% increase in F1 score (the weighted average of precision and recall) on the Southern California dataset. These advancements underscore the potential of array-based methods for more robust and generalizable phase picking.
However, it remains unclear whether a network-aware picker trained on earthquake data can be transferred efficiently to sparse, campaign-based microseismic arrays.

\begin{figure*}[htbp]
\centering
\includegraphics[width=0.75\textwidth]{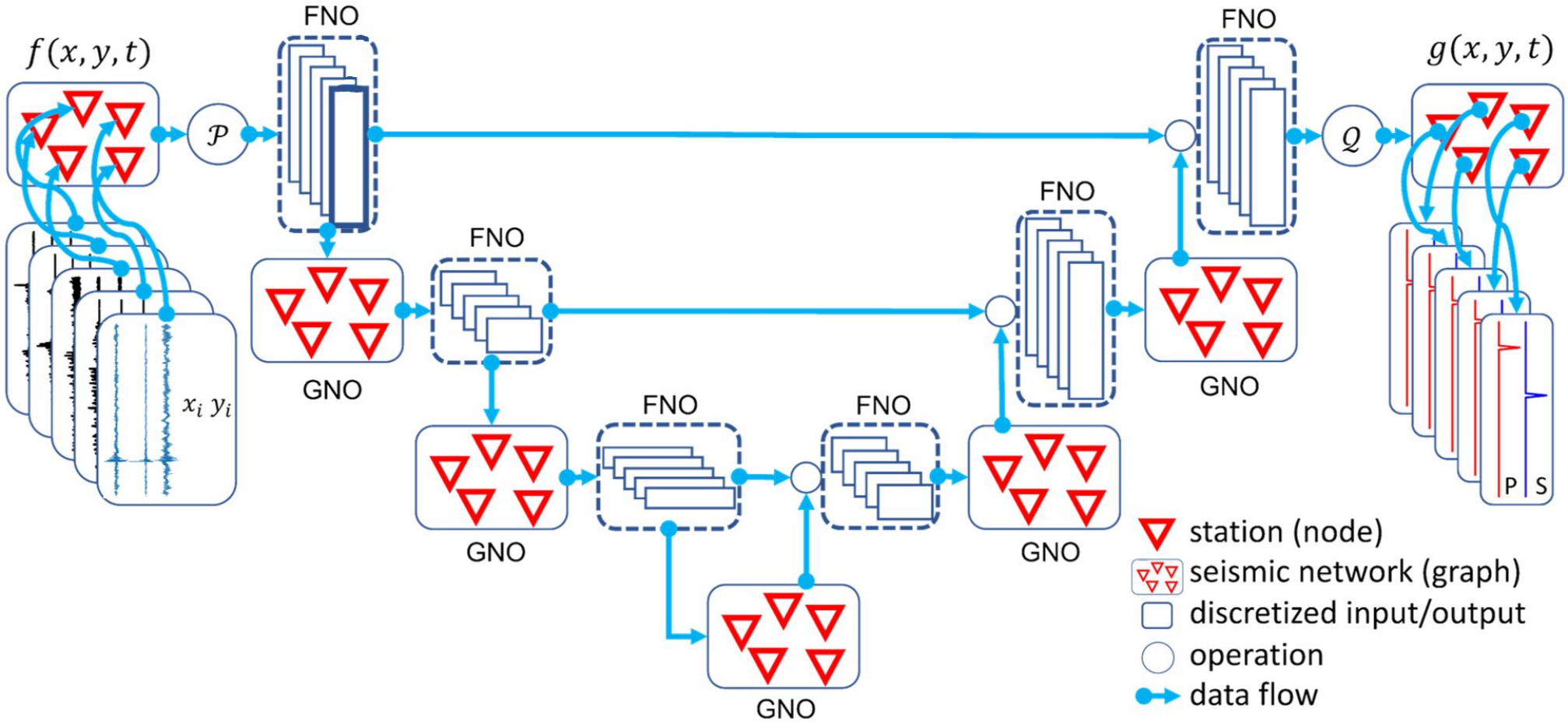}
\caption{
The architecture of the Phase Neural Operator. The network is built from multiple layers of Fourier Neural Operators (FNO) and Graph Neural Operators (GNO) arranged sequentially and iteratively. The up and down projections, denoted by P and Q, respectively, are implemented using neural networks. The model processes seismograms collected from a seismic network with arbitrarily located stations and predicts the P-phase and S-phase arrival time probabilities for each station. In addition to the three channels containing the three-component waveforms, the station locations are encoded as two separate channels, and the relative coordinates ($x_i$, $y_i$) between stations are used to learn the edge weights in the graph (adapted after Sun et al. (2023)~\cite{sun2023phase}).
}
\label{phaseno_scheme}
\end{figure*}

We address this question by adapting the Phase Neural Operator (PhaseNO)~\cite{sun2023phase} (Figure~\ref{phaseno_scheme}), a network-wide phase-picking algorithm based on the advanced Neural Operators framework~\cite{kovachki2023neural}. Designed to work with functions rather than finite-dimensional vectors, the neural operators learn mappings between function spaces and have become powerful tools for modeling partial differential equations (PDEs) and in other scientific applications. By incorporating spatio-temporal context, PhaseNO is capable of simultaneously picking seismic phases across arbitrary network geometries. Hence, it has the potential to solve one of the major challenges of microseismicity: detecting phase arrivals on variable non-stationary networks. Sun et al. (2023) applied the PhaseNO to a stationary network~\cite{sun2023phase}; we show how this methodology can be extended for new local networks with transfer learning.

We therefore evaluate PhaseNO on three real microseismic datasets using both full-network fine-tuning and a parameter-efficient CR-LoRA adaptation. In addition to comparing the fine-tuned model with the original pre-trained PhaseNO and a conventional STA/LTA-based picking workflow, we systematically benchmark its performance against two widely used single-station deep learning models, PhaseNet and EQTransformer. By leveraging spatio-temporal context across the entire seismic network, the fine-tuned PhaseNO consistently achieves better cross-dataset generalization and improved alignment with analyst picks. The central question of this study is whether a PhaseNO model trained on regional earthquake data can be adapted to microseismic data while updating only a small subset of parameters. By introducing and evaluating a parameter-efficient CR-LoRA strategy, we demonstrate that global spatio-temporal representations learned from large-scale tectonic datasets remain reusable for sparse, campaign-based microseismic deployments, requiring updates to only a small fraction of model parameters.

\section*{Data}

\begin{figure}[!ht]
  \centering
    \includegraphics[width=0.7\linewidth]{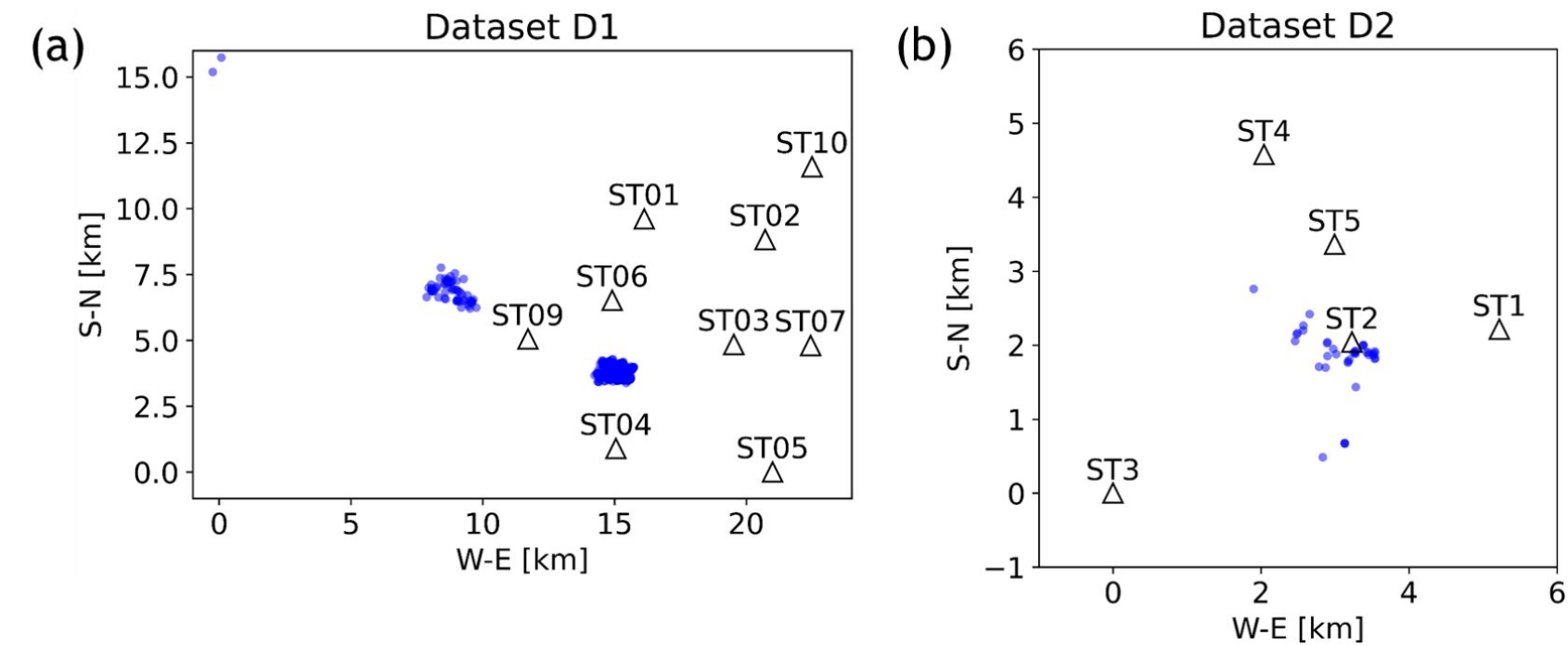}%
        \caption{
            The maps illustrate the locations of datasets D1 (a) and D2 (b) with seismic stations and induced event epicenters. Network stations are marked with triangles. The events occur at an average depth of 2.1 km, ranging from 1.7 to 2.4 km (D1), and 3.1 km, ranging from 2.8 to 3.4 km (D2).
            }
  \label{data_maps}
\end{figure}

This study utilizes three real microseismic datasets, each characterized by significantly different monitoring array designs and seismicity features (Figure~\ref{data_maps}). The first dataset, D1, consists of approximately 600 triggered microseismic events recorded during hydraulic fracturing operations in British Columbia (Canada), along with manually picked P- and S-wave phase arrival times and the seismic catalog. The data were collected using nine three-component surface seismic sensors with a sampling rate of 250 Hz, deployed on a surface area of approximately 100 km$^2$. Negative samples were extracted from background-noise windows in continuous recordings during intervals with no detected seismicity before operations began.

The second dataset D2 consists of 39 triggered events recorded during hydraulic fracturing operations in Ohio (United States). The induced seismicity data were acquired with five three-component surface seismic stations with a sampling rate of 250 Hz, covering an area of approximately 25 km$^2$. As with D1, D2 has manually picked phase arrival times and an accompanying location catalog.

The third dataset (D3) comprises 20 triggered records acquired during industrial operations in the United States. The monitoring array consisted of eleven three-component surface stations sampling at 500 Hz, deployed across an area of approximately 800 km$^2$. The dataset includes an equal proportion of local seismic events and false triggers, providing a balanced set for evaluating event discrimination performance.

\chreplaced{To ensure compatibility with the input dimensions required by the original PhaseNO model, seismograms from all the datasets were resampled to 100~Hz. Each triggered record spans approximately 60~seconds; from it, we extracted up to three 30-second windows (3000 samples each) by sliding the window so that the seismic arrivals appear near the beginning, the middle, and the end of the segment, respectively. The final utilized dataset comprised 1,460 waveform windows derived from approximately 650 labeled seismic triggers. These include mostly local microseismic and some near-field events (at least 10~km from the nearest station). Our data were partitioned into five sets:}{To ensure compatibility with the input dimensions required by the original PhaseNO model, seismograms from all the datasets were resampled to 100 Hz and divided into 30-second segments (each comprising 3000 samples). The final utilized dataset comprised 1,460 waveform segments derived from approximately 650 labeled seismic triggers. These include mostly local microseismic and some near-field events (at least 10 km from the nearest station). Our seismic events were partitioned into five sets:}

\begin{itemize}
    \item Training: 100 local and 100 noise-only \chreplaced{windows}{samples} from the 9-station dataset D1,
    \item Validation: \chreplaced{50 local windows}{50 local events} from D2 and 50 noise-only \chreplaced{windows}{samples} from D1, converted to five-station graphs by randomly selecting five D1 stations to match the D2 network size,
    \item Testing (3 sets): 
        \begin{itemize}
            \item[$\diamond$] 500 local and 500 noise-only \chreplaced{windows}{samples} from D1,
            \item[$\diamond$] \chreplaced{50 local windows}{50 local events} from D2 and 50 noise-only \chreplaced{windows}{samples} re-grouped from D1,
            \item[$\diamond$] 30 local and 30 false \chreplaced{trigger windows}{events} from the 11-station dataset D3.
        \end{itemize}
\end{itemize}

\noindent We ensured that no waveform segment was reused across training, validation, and test partitions for either event or noise samples. The training set was used to re-train the model, the validation set to adjust the learning rate and early stopping, and the test set to evaluate the performance. To account for labeling uncertainty, we used the probability density function in the shape of a triangle with a base of 0.4 sec, centered on the analyst picks (same function used in the original paper).

\section*{Methodology}

\subsection*{ Original Model }

We employ the Phase Neural Operator (PhaseNO) model~\cite{sun2023phase} as the core of our phase picking approach. The model uses two specialized neural operators to capture the intrinsic structure of seismic data. Temporal information is processed using layers of Fourier neural operators (FNO)~\cite{li2020fourier}, which use fast Fourier transforms to efficiently encode regularly sampled seismograms. Spatial information, on the other hand, is modeled using graph neural operators (GNO)~\cite{li2020neural}, which use message passing~\cite{gilmer2017neural} to aggregate features across irregularly distributed sensors. This dual approach ensures robust phase picking across varying seismic network configurations.

To further enhance performance on microseismic data, we developed a transfer learning (TL) workflow based on a pre-trained PhaseNO model originally trained on over 57,000 three-component seismic records from natural earthquakes in Northern California (1984–2019) and noise waveforms from the STanford EArthquake Dataset (STEAD). Although the pre-trained model was designed for seismograms with source-receiver distances ranging from a few to tens of kilometers, it demonstrated reasonable performance on our microseismic datasets. Our objective was not to further improve the original earthquake PhaseNO model, but to transfer its learned, network-level representation to a microseismic regime with scarce labeled data. The fine-tuning step is intended to adapt the pre-trained model to microseismic picking conventions (peaks/troughs), source receiver distances, stations placed in sedimentary basins and survey-specific noise characteristics, while preserving the information extracted from abundant earthquake datasets. We fine-tuned PhaseNO using a subset of our data that meets the original training requirements, i.e., normalized three-component seismograms with reference P- and S-wave arrival times.

\chadded{Because our adaptation builds directly upon the original PhaseNO architecture~\cite{sun2023phase}, the backbone hyperparameters are inherited unchanged. For completeness and reproducibility, Table~\ref{tab:phaseno_hyperparams} summarizes the key architectural parameters of the base model, including the arrangement of the 8 FNO layers, hidden channel dimensions, Fourier modes, and the GNO message-passing aggregation method. The full implementation details are available in the open-source repositories associated with the original study and our modified CR-LoRA framework.}

\begin{table}[htbp]
\centering
\caption{\chadded{Key hyperparameters of the inherited base PhaseNO architecture~\cite{sun2023phase}.}}
\label{tab:phaseno_hyperparams}
\begin{tabular}{ll}
\hline
\textbf{Parameter} & \textbf{Value} \\
\hline
Number of FNO layers & 8 \\
Number of GNO layers & 6 \\
Base hidden dimension ($w$) & 48 \\
Base Fourier modes ($m$) & 24 \\
Hidden dimension multipliers & $1\times, 2\times, 4\times, 8\times$ (U-shaped) \\
Fourier modes multipliers & $1\times, 1/2\times, 1/3\times$ \\
GNO message-passing aggregation & Sum \\
Skip connections & Concatenation \\
\hline
\end{tabular}
\end{table}

\subsection*{ Regular Fine-tuning }

\begin{figure}[!ht]
\centering
\includegraphics[width=0.7\linewidth]{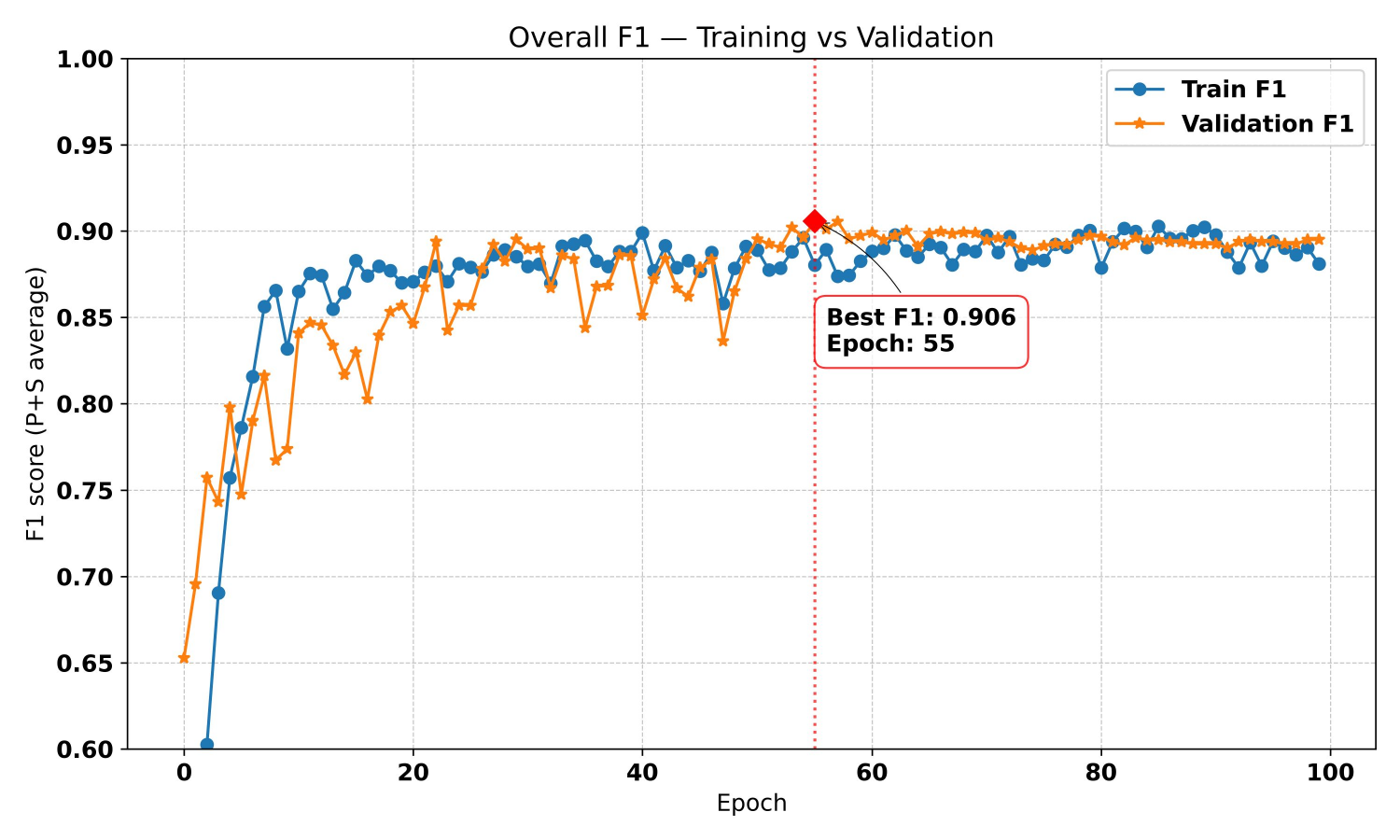}
\caption{
Evolution of the average P– and S–phase F1 score during training. Shown are the training and validation F1 curves over 100 epochs, with the validation performance peaking at an F1 score of 0.906 at epoch 55.
}
\label{f1_loss_curve}
\end{figure}

During TL, we retained the overall architecture of PhaseNO and initialized the weights with the pre-trained parameters. The fine-tuning process was conducted using the AdamW optimizer with a learning rate of 5e-5, a batch size of 1, and the ReduceLROnPlateau scheduler.
Model parameters were optimized using the BCEWithLogitsLoss function. The average P/S F1 score on the validation set was used for model selection and early stopping. Training was stopped if the validation F1 score did not improve for 10 consecutive epochs (Figure~\ref{f1_loss_curve}).

\begin{figure}[!ht]
\centering
\includegraphics[width=0.7\linewidth]{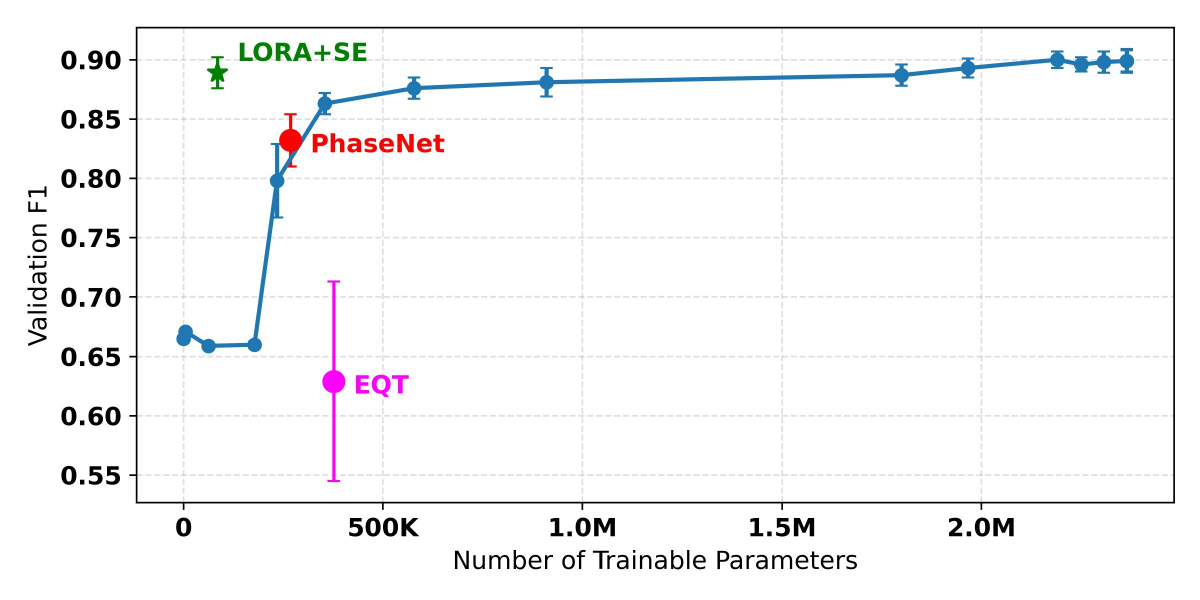}
\caption{
Impact of progressive unfreezing of PhaseNO layers on validation F1 score as a function of the number of trainable parameters. Blue markers show the mean validation F1 ($\pm 1$ SD) over five independent fine-tuning runs for the original PhaseNO backbone with progressively unfrozen FNO/GNO blocks. Full-network fine-tuning (>2.0M trainable parameters) approaches a validation F1 of 0.90.
The green star denotes the proposed LoRA+SE (CR-LoRA) adaptation, which achieves a validation F1 of $0.889 \pm 0.013$ using only 85K trainable parameters (3.6\% of the full model). Despite its parameter efficiency, the LoRA+SE model attains performance comparable to extensive backbone unfreezing. For comparison, PhaseNet (red circle; 268K parameters) yields $0.832 \pm 0.022$, whereas EQTransformer (EQT) (magenta circle; 377K parameters) achieves $0.629 \pm 0.084$ under similar fine-tuning conditions.
}
\label{valF1_vs_unfreeze}
\end{figure}

We performed a cross-validation study with the same model hyperparameters only varying a set of trainable PhaseNO blocks (Figure~\ref{valF1_vs_unfreeze}). Each case involved 5 fine-tuning runs starting with the pre-trained weights. The study showed that the maximum validation F1 scores can be achieved with unfreezing more PhaseNO blocks, and the best model was chosen from the run with all the layers trainable. Training from scratch with the same model architecture but starting with random weights did not reach even 20\% for the F1 score. We then evaluated the performance of the fine-tuned model by comparing the predicted arrival times with reference labels, using key confusion matrix metrics to comprehensively assess detection performance.

\subsection*{ Parameter-efficient Fine-tuning }

In addition to conventional full-network fine-tuning via progressive layer unfreezing, we developed a parameter-efficient adaptation strategy tailored to microseismic data. The proposed approach augments the pre-trained PhaseNO architecture with Low-Rank Adaptation (LoRA) modules and Squeeze-and-Excitation (SE) channel-recalibration blocks, resulting in a Channel-Recalibrated Low-Rank Adaptation (CR-LoRA) variant of PhaseNO.

The CR-LoRA model freezes the pre-trained backbone and injects low-rank adapters into (i) the pointwise 1D convolutions within the FNO blocks and (ii) the linear layers of the GNO feedforward modules. Each LoRA module decomposes the weight update into two small trainable matrices of rank $r = 4$, scaled by $\alpha/r$, ensuring that the initial forward pass remains identical to the pre-trained model. This design preserves the global spatio-temporal representation learned from large earthquake datasets while enabling controlled domain adaptation.

To address the strong variability of noise and amplitude characteristics in microseismic campaigns, we further introduce a 1D Squeeze-and-Excitation (SE) block \chreplaced{that exclusively recalibrates the LoRA branch output before it is merged with the base convolution}{after each pointwise convolution}. The SE mechanism performs global average pooling along the temporal dimension, followed by a bottleneck fully connected transformation and sigmoid gating to dynamically reweight feature channels. While LoRA captures low-dimensional domain shifts, the SE blocks adaptively recalibrate channel importance in response to survey-specific noise profiles. 
\chreplaced{This design ensures that the SE mechanism focuses purely on recalibrating the newly learned, domain-specific adaptations without altering the robust, frozen pre-trained features. An ablation study comparing CR-LoRA, LoRA-only, and SE-only variants is presented in the Results section.}{This integration has been found to be critical for the stable adaptation to low SNR, peak/trough-labeled microseismic data, for which the use of traditional low-rank adaptation alone did not yield the desired convergence.}

For the model, only the LoRA parameters, the SE blocks, and the decision head (fc1 and fc2) are trainable, with the remaining weights remaining frozen. The total number of trainable parameters is approximately 85,000, which is only 3.6\% of the total PhaseNO model, which has approximately 2.4 million parameters. This significant reduction acts as an implicit regularizer and prevents overfitting on the relatively small microseismic calibration dataset, which has 200 segments.

Since only a small percentage of the total number of parameters is being optimized, the training strategy has been adjusted. Unlike the full-network fine-tuning strategy, which involved the use of progressive unfreezing of the network’s layers and the use of stronger weight decay, the adapter-based training strategy used AdamW optimizer with lower weight decay, which aligns with the intrinsic regularization induced by the low-rank constraint. Instead of the ReduceLROnPlateau learning rate scheduler used with the regular fine-tuning strategy, we adopted the Cosine Annealing over 100 epochs, which enables the learning rate to decay smoothly and proactively toward its minimum value. This has been found to be effective in avoiding the shallow local minima, which are often encountered during the optimization of the small adapter modules.

\subsection*{ Evaluation Metrics }

Parameters that are derived from confusion matrix are precision, recall, F1-score, and accuracy. Precision measures the proportion of correct detections to all detections that are identified, whereas recall measures the ability of the model to identify known phase arrivals. F1-score is defined as the harmonic mean of precision and recall, thus balancing both false positives and false negatives. Accuracy is calculated as the proportion of correctly identified to all instances.
Together, these metrics provide an overall assessment of detector performance, balancing the need to minimize false alarms and maximize the detection of correct phase arrivals. Note, the confusion matrix derived metrics do not account for errors in the manually labeled data. The number of true positives, false positives, false negatives, and true negatives was computed for each station and then summed for all events to get the final scores. The arrival time difference between manual and model picks was considered correct if it was less than the chosen threshold of 0.5 sec (same with the original PhaseNO paper).

\section*{Results and Discussion}

By transfer learning, we overcome the difference in spatial scale between our microseismic data and the natural earthquake data used for original training. Parameter-efficient fine-tuning the pre-trained model required only 200 seismograms (200 / 57,000 = 0.4\% of the original training dataset), demonstrating the efficiency of this approach. The CR-LoRA PhaseNO model achieved close agreement with the analyst-provided labels while substantially reducing inference time \chadded{compared to manual analytical workflows (achieving an average processing time of 10.0~ms per 5-station event on a single NVIDIA RTX PRO 6000 GPU)}.
 
\begin{figure*}[htbp]
\centering
\includegraphics[width=0.7\linewidth]{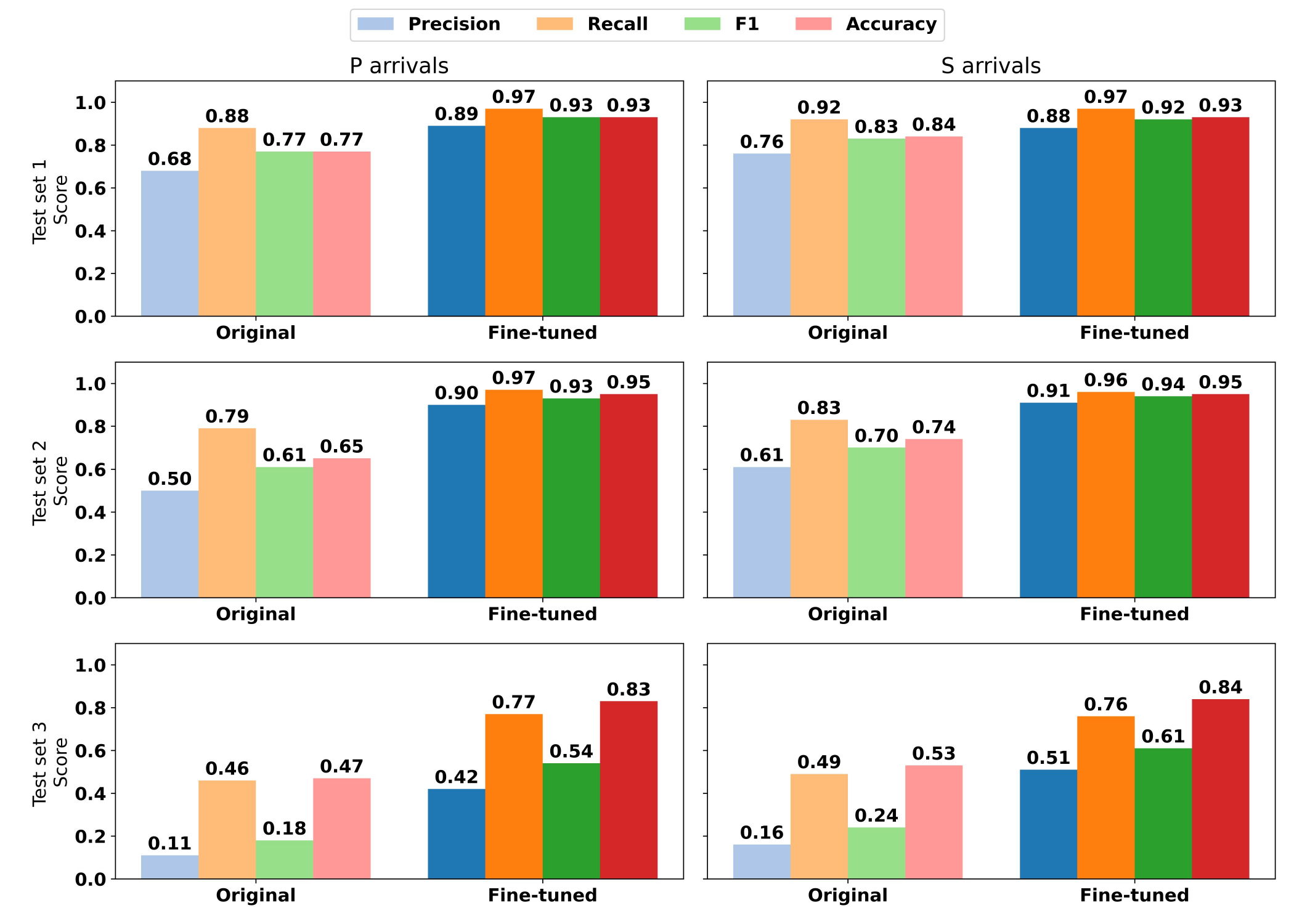}
\caption{
A comparison of performance between the original PhaseNO and the fine-tuned model for P and S arrival picks for the three test sets. We use precision, recall, F1, and accuracy scores to quantify and compare the performance.
}
\label{comp_3tests}
\end{figure*}

Quantitative evaluation using confusion matrix metrics further underscored the improvements brought by fine-tuning. For test set \#1, the fine-tuned model demonstrated up to a 12-21\% increase in precision, 5-9\% in recall, 9-16\% in F1 score, and 9-16\% in overall accuracy relative to the original PhaseNO (Figure~\ref{comp_3tests}). The original model retained relatively high recall but produced substantially more false positives, which reduced precision and F1 score before fine-tuning.

Similar improvements were observed in dataset D2 (Figure~\ref{comp_3tests}), which included a more sparse set (five stations compared to nine in D1), highlighting the ability to use the fine-tuned model on a new network, which was not used for training. This advantage is of high importance for induced seismicity as it allows the use of sparse networks without prior induced seismicity detected on these networks. On D3, absolute performance decreased for all methods, but the fine-tuned model still improved F1 score over the original PhaseNO by more than 30\%.

\chadded{To ensure these performance improvements were not attributable to random statistical fluctuations, given the relatively small number of samples in the test sets, we conducted a non-parametric Wilcoxon signed-rank test comparing the per-event F1 scores of the original and fine-tuned models. The tests yielded highly significant results ($p < 0.0001$) across all three datasets (D1 $N=502$, D2 $N=50$, and D3 $N=33$ valid events), confirming that the improvements afforded by the CR-LoRA adaptation are statistically robust despite the small sample sizes in D2 and D3.}

To provide a benchmark against which to evaluate the machine-learning results, we implemented a classical picking workflow combining STA/LTA triggering and polarization analysis. P-wave arrivals were first identified on the vertical component using a short-term/long-term average (STA/LTA) detector, where exceedance of a predefined onset threshold marked the preliminary pick. This initial estimate was subsequently refined within a ±0.5-s window using the Akaike Information Criterion (AIC) to obtain a more stable onset time. S-wave picking was performed on the three-component seismograms by scanning a sliding window and locating the time at which the horizontal-to-total energy ratio is maximized, constrained to occur after the P arrival and within a physically reasonable S–P interval. This hybrid STA/LTA-AIC P Picker combined with the polarization-based S Picker is a conventional signal-processing method to compare against in local earthquake analysis. 

\chadded{Instead of performing an exhaustive hyperparameter grid-search for the conventional baseline, we established the configuration based on empirical conventions for high-frequency local microseismic monitoring. These settings are engineered to handle the impulsive, short-duration characteristics of microseismic events while filtering out standard environmental noise (e.g., oceanic microseisms below 1~Hz and cultural noise above 40~Hz). Table~\ref{tab:conventional_params} summarizes the selected parameters for the pre-processing bandpass filter, the STA/LTA-AIC P-wave picker, and the polarization-based S-wave picker. These parameters remain fixed across all testing scenarios, which reflects standard operational practices where continuous tuning per dataset is not feasible.}

\begin{table*}[htbp]
\centering
\caption{\chadded{Parameter configuration for the conventional picking baseline.}}
\label{tab:conventional_params}
\begin{tabular}{@{}ll@{}}
\toprule
\textbf{Processing Stage} & \textbf{Parameter Settings} \\
\midrule
\textbf{1. Pre-Processing} & Filter Type: 4th-order Butterworth bandpass \\
(All components) & Frequency Band: 2.0 -- 38.0 Hz \\
\midrule
\textbf{2. P-Wave Picking} & Component: Vertical (Z) \\
(STA/LTA + AIC) & STA Window: 0.5 s \\
& LTA Window: 5.0 s \\
& Trigger Threshold (ON): 3.0 \\
& AIC Refinement Window: $\pm$0.5 s \\
\midrule
\textbf{3. S-Wave Picking} & Component: 3-component (E, N, Z) \\
(Polarization Analysis) & Search Window Limits (S-P interval): 0.02 s to 5.0 s \\
& Sliding Window Length: 0.1 s \\
& Sliding Window Step Size: 0.01 s \\
& Minimum Horizontal-to-Total Energy Ratio: 0.5 \\
\bottomrule
\end{tabular}
\end{table*}

\begin{figure*}[htbp]
\centering
\includegraphics[width=0.7\linewidth]{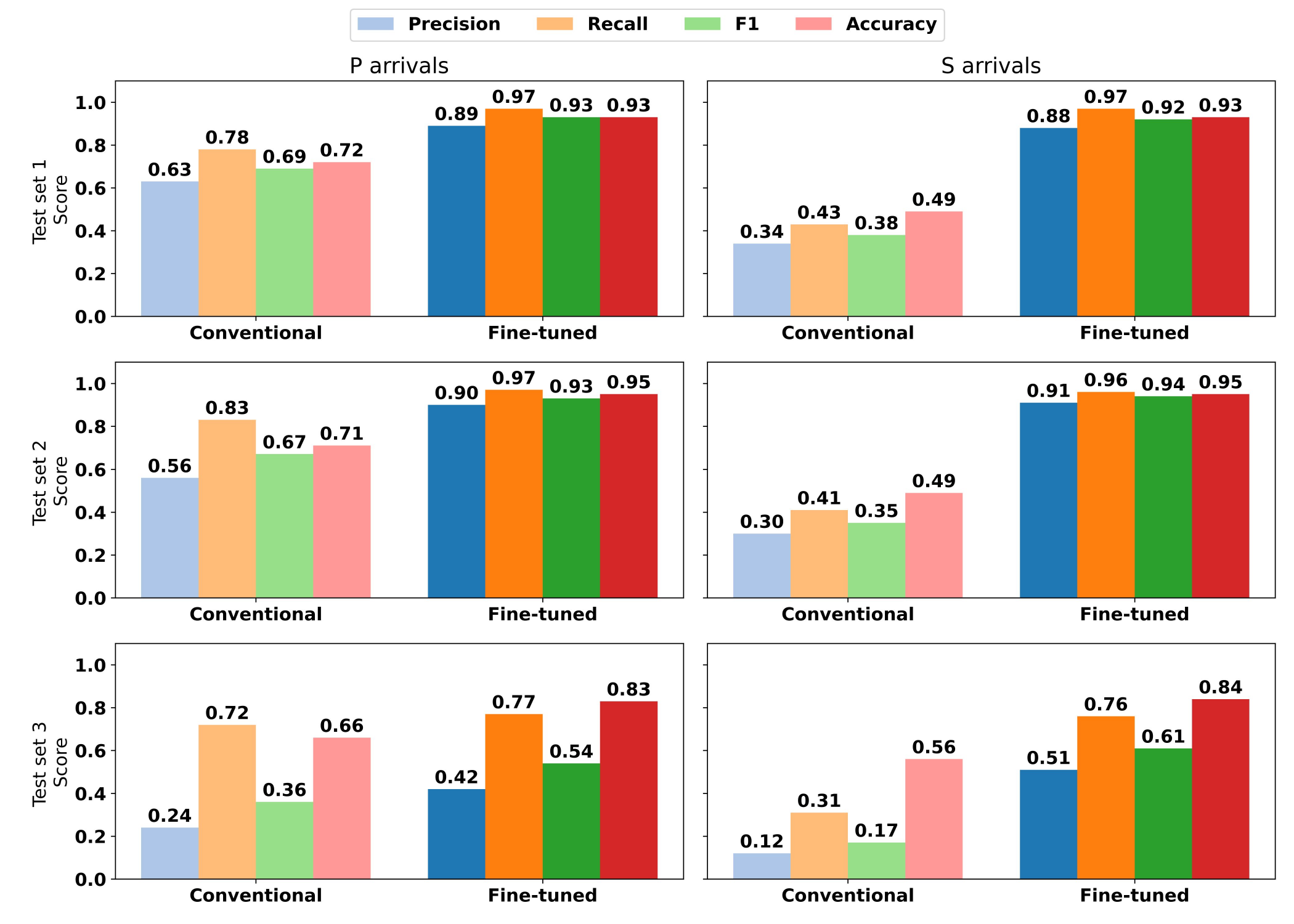}
\caption{
A comparison of performance between the conventional STA/LTA method and the fine-tuned model for P and S arrival picks for the three test sets. We use precision, recall, F1, and accuracy scores to quantify and compare the performance.
}
\label{comp_conv_3tests}
\end{figure*}

With reference to the three test sets considered in this work (Figure~\ref{comp_conv_3tests}), the fine-tuned model achieved higher confusion matrix metrics than the conventional method.

This study also compares CR-LoRA PhaseNO against two state-of-the-art baselines that are commonly utilized in a single-station deep learning setting, namely PhaseNet~\cite{zhu2019phasenet} and EQTransformer~\cite{mousavi2020earthquake} (see also Figure~\ref{valF1_vs_unfreeze}). PhaseNet and EQTransformer are capable of independent phase detection and picking at individual stations using CNNs and related modern deep learning components. PhaseNet is originally pre-trained on a large Northern California dataset consisting of 623,054 manually labeled picks of P and S phases. EQTransformer is originally pre-trained on a large STEAD dataset~\cite{mousavi2019stanford}. To compare against PhaseNO, PhaseNet and EQTransformer are fine-tuned on the same microseismic training, validation, and testing sets, and the comparison is provided in Table~\ref{tab:model_comparison}.

\begin{table*}[ht]
\centering
\caption{
Performance comparison of seismic phase picking models across three test sets. Metrics for P-arrival and S-arrival include Precision (Pr), Recall (Re), F1-score (F1), and Accuracy (Acc). The original and fine-tuned models are compared against Conventional methods, PhaseNet, and EQ Transformer.
}
\label{tab:model_comparison}
\begin{tabular}{@{} l cccc cccc @{}}
\toprule
& \multicolumn{4}{c}{\textbf{P-Arrival}} & \multicolumn{4}{c}{\textbf{S-Arrival}} \\
\cmidrule(lr){2-5} \cmidrule(l){6-9}
\textbf{Model} & \textbf{Pr} & \textbf{Re} & \textbf{F1} & \textbf{Acc} & \textbf{Pr} & \textbf{Re} & \textbf{F1} & \textbf{Acc} \\
\midrule
\multicolumn{9}{@{}l}{\textbf{Test Set 1}} \\
\midrule
Original       & 0.68 & 0.88 & 0.77 & 0.77 & 0.76 & 0.92 & 0.83 & 0.84 \\
Fine-tuned     & \textbf{0.89} & \textbf{0.97} & \textbf{0.93} & \textbf{0.93} & \textbf{0.88} & \textbf{0.97} & \textbf{0.92} & \textbf{0.93} \\
Conventional   & 0.63 & 0.78 & 0.69 & 0.72 & 0.34 & 0.43 & 0.38 & 0.49 \\
PhaseNet       & 0.89 & 0.28 & 0.43 & 0.66 & 0.88 & 0.30 & 0.45 & 0.67 \\
EQ Transformer & 0.74 & 0.51 & 0.61 & 0.70 & 0.75 & 0.54 & 0.63 & 0.72 \\
\midrule
\multicolumn{9}{@{}l}{\textbf{Test Set 2}} \\
\midrule
Original       & 0.50 & 0.79 & 0.61 & 0.65 & 0.61 & 0.83 & 0.70 & 0.74 \\
Fine-tuned     & 0.90 & \textbf{0.97} & \textbf{0.93} & \textbf{0.95} & 0.91 & \textbf{0.96} & \textbf{0.94} & \textbf{0.95} \\
Conventional   & 0.56 & 0.83 & 0.67 & 0.71 & 0.30 & 0.41 & 0.35 & 0.49 \\
PhaseNet       & \textbf{0.93} & 0.83 & 0.88 & 0.92 & \textbf{0.96} & 0.85 & 0.90 & 0.93 \\
EQ Transformer  & 0.76 & 0.74 & 0.75 & 0.82 & 0.79 & 0.75 & 0.77 & 0.82 \\
\midrule
\multicolumn{9}{@{}l}{\textbf{Test Set 3}} \\
\midrule
Original       & 0.11 & 0.46 & 0.18 & 0.47 & 0.16 & 0.49 & 0.24 & 0.53 \\
Fine-tuned     & 0.42 & \textbf{0.77} & 0.54 & 0.83 & \textbf{0.51} & \textbf{0.76} & \textbf{0.61} & \textbf{0.84} \\
Conventional   & 0.24 & 0.72 & 0.36 & 0.66 & 0.12 & 0.31 & 0.17 & 0.56 \\
PhaseNet       & \textbf{0.51} & 0.59 & \textbf{0.55} & \textbf{0.87} & 0.51 & 0.51 & 0.51 & 0.84 \\
EQ Transformer & 0.13 & 0.63 & 0.22 & 0.39 & 0.15 & 0.61 & 0.25 & 0.40 \\
\bottomrule
\end{tabular}
\end{table*}

PhaseNet shows high precision scores in all three test sets, especially in P arrivals (0.93 in Test Set 2 and 0.51 in Test Set 3). However, this is often at the expense of lower recall scores, as shown in Test Set 1, where PhaseNet reports lower recall scores of 0.28 (P arrivals) and 0.30 (S arrivals), resulting in lower F1 scores of 0.43 and 0.45, respectively. This shows that PhaseNet is able to detect arrivals with high confidence when it triggers but is unable to detect a large number of actual arrivals in more complex or low-SNR conditions. These results suggest that PhaseNet is able to perform well under stable conditions but is unable to generalize to heterogeneous microseismic deployments due to its single-station approach. 
EQTransformer has a better balance between precision and recall than PhaseNet, as shown in Test Sets 1 and 2, resulting in moderate F1 scores between 0.61 and 0.77. However, this is still lower than the fine-tuned PhaseNO model and shows more variability between runs (Figure~\ref{valF1_vs_unfreeze}), indicating sensitivity to noise and lower robustness in sparse network configurations. The results show that although PhaseNet and EQTransformer are strong baselines, they are still unable to match the stability and cross-dataset consistency achieved by the network-wide, fine-tuned PhaseNO approach.

\chadded{To empirically validate the design of the CR-LoRA adaptation strategy, we conducted an ablation study comparing three model variants: (1)~CR-LoRA, which combines LoRA adapters with SE channel recalibration applied exclusively to the LoRA branch; (2)~LoRA-only, which removes the SE blocks entirely; and (3)~SE-only, which removes the LoRA adapters and retains only the SE blocks and the trainable decision head. All three variants were evaluated under an identical 5-fold cross-validation protocol, and the single best checkpoint (selected by validation F1) for each variant was then evaluated on the three test datasets.}

\begin{table*}[ht]
\centering
\caption{\chadded{Ablation study: 5-fold cross-validation F1 scores. The best fold for each variant (used for test evaluation) is shown in bold.}}
\label{tab:ablation_cv}
\begin{tabular}{@{} l ccccc cc @{}}
\toprule
\textbf{Model Variant} & \textbf{F0} & \textbf{F1} & \textbf{F2} & \textbf{F3} & \textbf{F4} & \textbf{Mean} & \textbf{SD} \\
\midrule
CR-LoRA    & 0.884 & 0.891 & 0.884 & \textbf{0.896} & 0.884 & \textbf{0.888} & \textbf{0.005} \\
LoRA-only  & 0.890 & 0.881 & \textbf{0.897} & 0.869 & 0.883 & 0.884 & 0.009 \\
SE-only    & 0.845 & \textbf{0.849} & 0.842 & 0.812 & 0.831 & 0.836 & 0.013 \\
\bottomrule
\end{tabular}
\end{table*}

\begin{table}[ht]
\centering
\caption{\chadded{Ablation study: test-set F1 scores (mean of P-wave and S-wave F1) for the best checkpoint of each variant. Bold values indicate the best score for each test set.}}
\label{tab:ablation_test}
\begin{tabular}{@{} l ccc @{}}
\toprule
\textbf{Model Variant} & \textbf{D1} & \textbf{D2} & \textbf{D3} \\
\midrule
CR-LoRA    & 0.929 & \textbf{0.930} & \textbf{0.594} \\
LoRA-only  & \textbf{0.929} & 0.928 & 0.589 \\
SE-only    & 0.910 & 0.870 & 0.537 \\
\bottomrule
\end{tabular}
\end{table}

\chadded{Tables~\ref{tab:ablation_cv} and~\ref{tab:ablation_test} summarize the ablation results. CR-LoRA achieves the highest mean validation F1 (0.888) with the lowest standard deviation (0.005), indicating the most stable optimization across folds. On the test datasets, CR-LoRA matches LoRA-only on the in-distribution D1 set, but outperforms it on the out-of-distribution D2 and D3 deployments, demonstrating improved generalization to unseen network geometries. The LoRA-only variant remains competitive, confirming that LoRA carries out the primary domain adaptation. By contrast, the SE-only variant significantly underperforms across all evaluations, with the lowest validation mean (0.836), the highest variance, and substantially degraded test-set generalization, particularly on D3 (0.537 F1). These results confirm that LoRA is essential for bridging the domain gap between earthquake and microseismic data, whereas the SE module alone lacks sufficient capacity for adaptation. The combined CR-LoRA design provides a complementary benefit: LoRA captures the low-rank domain shift, while the SE block recalibrates channel importance in a way that measurably improves robustness on sparse and dense deployments.}

\begin{figure*}[htbp]
\centering
\includegraphics[width=0.7\linewidth]{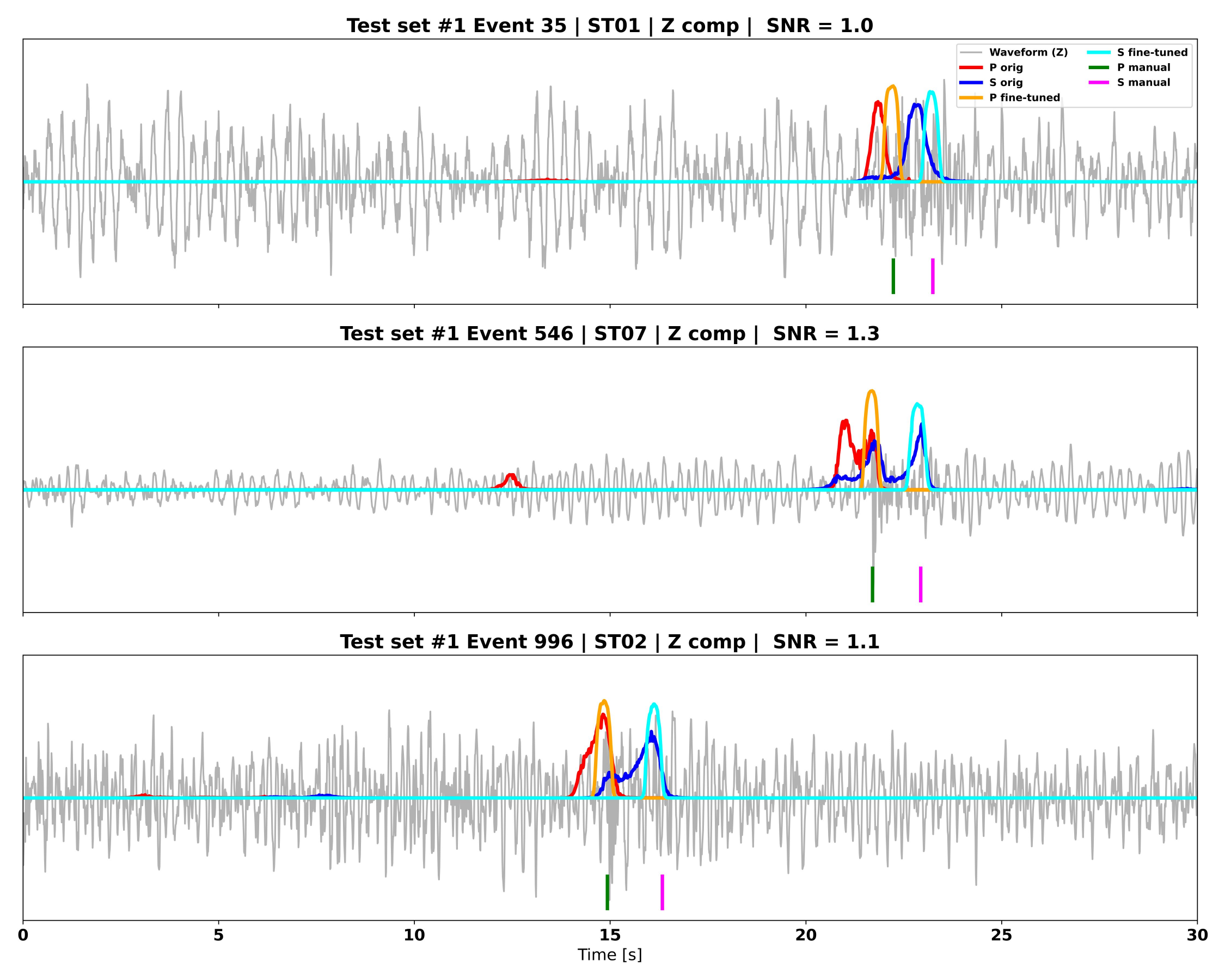}
\caption{
Example waveforms from test set \#1 demonstrate the probability results for the original (red and blue graphs) and fine-tuned (orange and cyan graphs) PhaseNO models with different signal-to-noise ratios (SNR). The red and orange graphs represent the P-pick probabilities, whereas the blue and cyan graphs represent the S-pick probabilities. The green and magenta vertical bars represent the analyst picks for the P and S waves, respectively. The results for the fine-tuned model are more accurate and closer to the analyst picks for low SNR values, thereby highlighting the robustness and accuracy achieved through the fine-tuning process.
}
\label{waveforms_snr}
\end{figure*}

Both the original and fine-tuned models were evaluated on a comprehensive test set comprising seismograms that were not used during training. Randomly selected waveforms and their corresponding picks from the test dataset show a strong agreement between the TL results and the analyst picks, even in low SNR conditions (Figure~\ref{waveforms_snr}).

\begin{figure*}[htbp]
\centering
\includegraphics[width=0.99\textwidth]{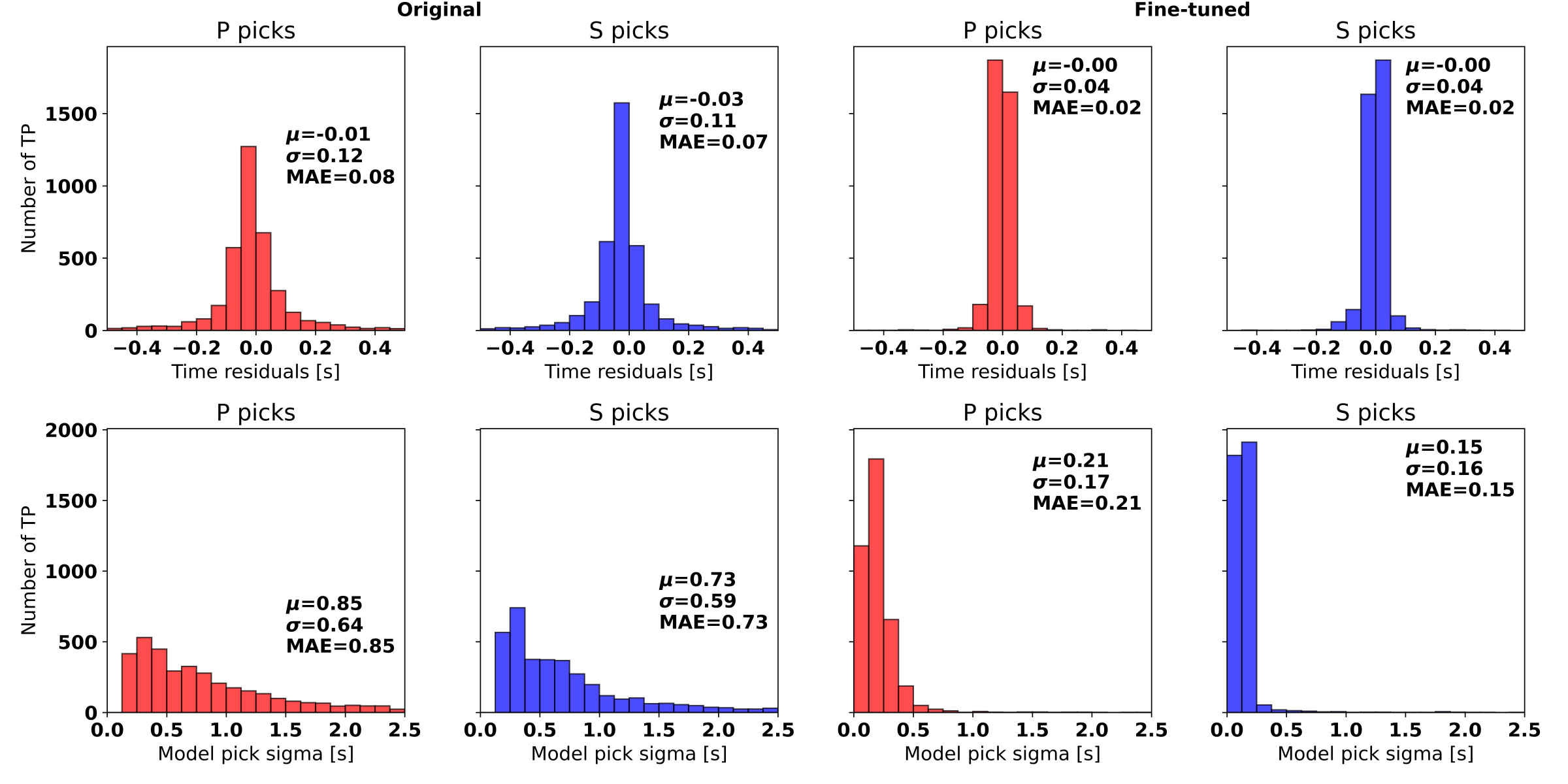}
\caption{
Comparison of differences in arrival times for picks (top row) and the calculated prediction uncertainty based on the probability function of each pick (bottom row) on test set \#1 of two deep learning models: the original PhaseNO (left) and the fine-tuned PhaseNO (right). The top row of plots shows a histogram of time residuals of P-wave picks (red) and S-wave picks (blue), i.e., differences between model predictions and analyst picks. Each plot provides information about the mean ($\mu$), standard deviation ($\sigma$), and mean absolute error (MAE) of all true positive picks with a time residual below 0.5 s. Note that the fine-tuned model has the mean residual closer to zero, and its average uncertainty is roughly three times smaller compared to the original PhaseNO, showing improved overall picking accuracy.
}
\label{TP_test1}
\end{figure*}

\begin{figure*}[htbp]
\centering
\includegraphics[width=0.99\textwidth]{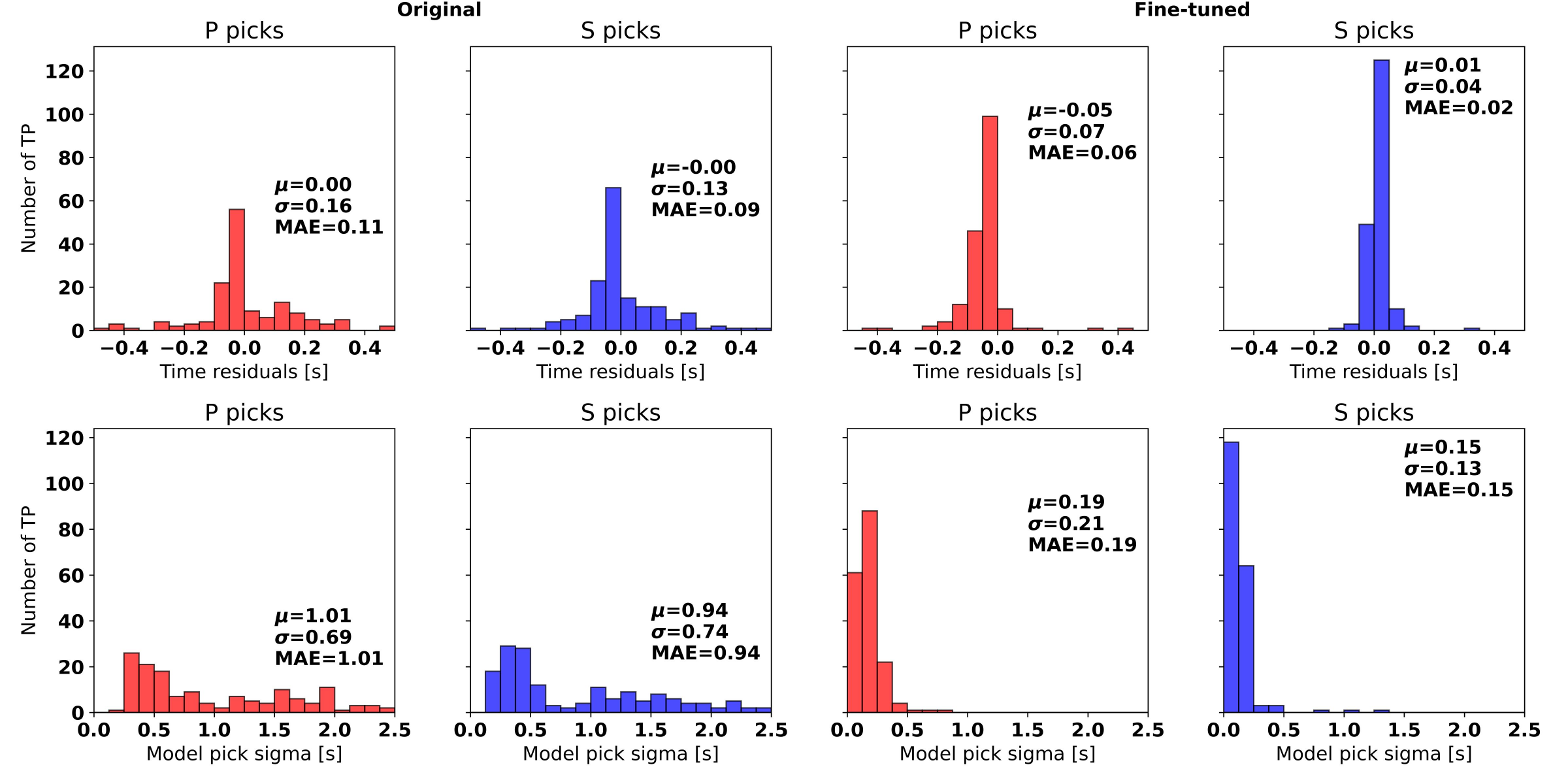}
\caption{
Comparison of picking arrival time differences (top row) and predicted uncertainty computed from the probability function of a corresponding pick (bottom row) on the test set \#2 for two deep-learning models: the original PhaseNO (left) and the fine-tuned PhaseNO (right). The top-row histograms show time residuals for P-wave picks (red) and S-wave picks (blue), defined as the difference between the model-predicted arrivals and the analyst picks. Each panel reports the mean ($\mu$), standard deviation ($\sigma$), and mean absolute error (MAE) for all true-positive (TP) picks with time residuals below 0.5 s. The bottom-row histograms display the distribution of 1-sigma probability intervals for TPs in each model. Similar to the D1 dataset performance, the fine-tuned model’s mean residual is closer to zero, and its average uncertainty interval is about three times smaller than that of the original PhaseNO, indicating improved overall picking accuracy.
}
\label{TP_test2}
\end{figure*}

Additionally, analysis of picking errors and predicted uncertainty intervals (Figure~\ref{TP_test1}) revealed that the original model picks have more negative bias in the difference between manual and model picks, and the fine-tuned picks do not have such bias. This observation probably results from the fact that the original earthquake training dataset was picked on the onset of arrivals, while the microseismic datasets were picked on peaks/troughs of the arrivals. The second important observation is the smaller uncertainty in the fine-tuned model pick times, resulting in a narrower probability distribution around the correct time (Figures~\ref{TP_test1} and~\ref{TP_test2}). The original model’s one-sigma intervals are approximately three times wider than those of the fine-tuned model, indicating that fine-tuning produces sharper and more confident picks. These results suggest that the fine-tuned model is not simply “more accurate” in an absolute sense; rather, it has transitioned from an onset-based earthquake labeling convention to a peak/trough-based microseismic convention, thereby making it more similar to the actual definition of phase arrival times as they are represented within microseismic data.

The analysis also considers how model performance changes based on adjustments to the true-positive time-tolerance threshold. \chreplaced{Figure 10 shows that as this threshold increases from a strict 0.1~s tolerance up to a relaxed 1.0~s tolerance in 0.1~s increments, the F1 score of the original PhaseNO model remains consistently lower than that of the fine-tuned model across the entire range.}{Figure 10 shows that as this threshold increases from 0.1 to 1.0 s, the F1 score of the original PhaseNO model is still lower than that of the fine-tuned model.}

\begin{figure*}[htbp] 
\centering
\includegraphics[width=0.9\linewidth]{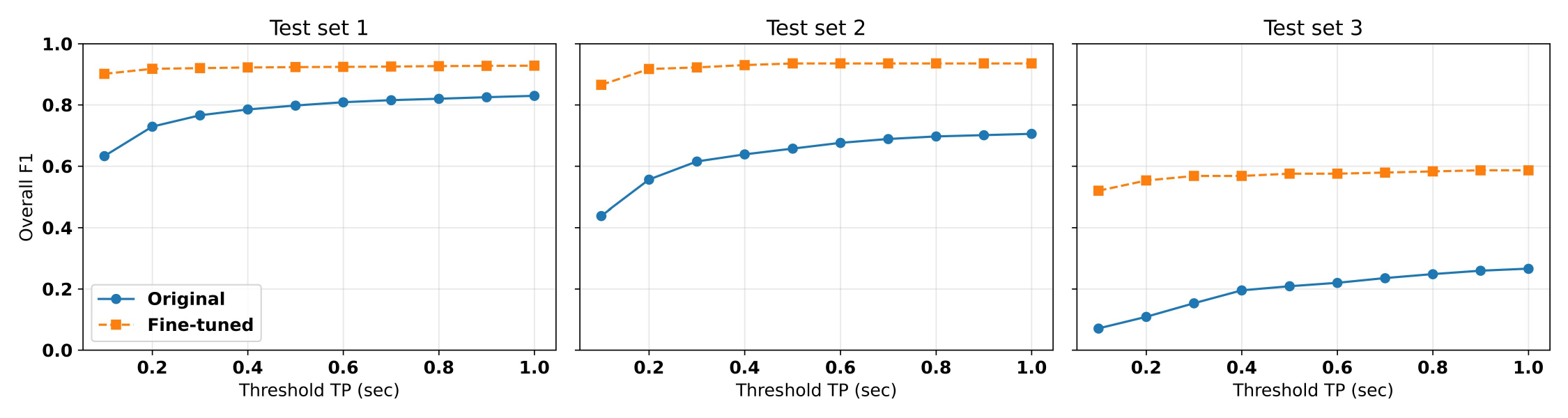}
\caption{
Sensitivity of the overall F1 score with respect to the true-positive time-tolerance threshold for both the original and fine-tuned models over the three test sets. \chadded{The evaluation ranges from a strict tolerance of 0.1~s to a relaxed 1.0~s, evaluated in 0.1~s increments.} For all test data, it is evident that the fine-tuned model significantly outperforms the original PhaseNO over the entire range \chreplaced{}{(from 0.1 to 1.0 seconds)}. Although increasing the tolerance gradually increases the F1 score for both models, the fine-tuned model performs significantly better \chadded{even under the strictest 0.1~s criterion}.
}
\label{f1_vs_thrTP}
\end{figure*}

Although this current study is specifically concerned with phase-level evaluation under benchmark conditions, it is evident that this adapted PhaseNO model has significant promise to improve precision-recall balance and reduce timing bias in association with continuous monitoring workflows. The ultimate evaluation of this model’s ability to improve catalog generation and association under continuous monitoring conditions is reserved for future work, as it necessitates additional design considerations beyond those presented here.

\section*{Conclusions}

Accurate phase picking is essential for seismic event detection and subsurface characterization. In particular, for induced microseismic monitoring, the signal-to-noise ratio is low, and acquisition geometries are different for each survey.

In this study, we propose and validate a transfer-learning approach for microseismic phase picking using the pre-trained Phase Neural Operator (PhaseNO), which has been previously trained on more than 57,000 three-component earthquake and noise records. We fine-tune the pre-trained model using only 200 microseismic and noise-only samples, corresponding to approximately 0.4\% of the original model's training set. The fine-tuning approach involves using a parameter-efficient adaptation mechanism, which only updates 3.6\% of the model's parameters. The quantitative analysis of performance compared to the number of trainable model parameters demonstrates that the majority of the model's weights are responsible for encoding the transferable spatio-temporal structure and not the survey-specific features.

Over three independent test sets with different network architecture and acquisition parameter setups, the fine-tuned model achieves up to 30\% F1 score improvement over the original PhaseNO, while also outperforming the STA/LTA AIC approach. In comparison with other deep-learning models such as PhaseNet and EQTransformer, also fine-tuned on the same conditions, the proposed network-wide model demonstrates higher stability and cross-dataset generality, especially for low-SNR and sparse network conditions. While the single-station models achieve high precision for certain conditions, their recall and robustness are compromised for more heterogeneous setups. In contrast, the Neural Operator approach leverages the spatial coherence among multiple stations more effectively and yields more balanced precision-recall performance and less timing bias.

Overall, the results demonstrate that large publicly available datasets and pre-trained network-wide Neural Operator models are viable for providing a robust solution for microseismic monitoring even with limited available calibration data. \chadded{From an engineering perspective, the CR-LoRA strategy demonstrates high deployment efficiency by updating only $\sim$85K trainable parameters (3.5\% of the 2.4M parameter backbone) and requiring a modest GPU memory footprint of 279~MB. In terms of inference speed, our array-based PhaseNO achieves an average GPU processing time of 10.0~ms per 5-station event on a single NVIDIA RTX PRO 6000. While this is higher than single-station models like PhaseNet (0.26~ms) or EQTransformer (5.2~ms) due to the overhead of cross-station message passing, it remains well within the sub-second requirements for real-time monitoring and historical archive processing.} The proposed MicroPhaseNO workflow is shown to provide a scalable and cost-effective solution to deploying advanced deep-learning models in data-constrained induced seismicity applications.

\section*{Declaration of generative AI and AI-assisted technologies in the manuscript preparation process}

\chreplaced{During the preparation of this work, the authors used the Gemini agent within the Google Antigravity IDE to assist with writing and troubleshooting code for data visualization and computational workflows, as well as to improve the readability and language of the manuscript. The authors reviewed and edited the output as needed and take full responsibility for the content of the published article.
}{
During the preparation of this work, the authors used ChatGPT (OpenAI, USA) to refine the language and improve the readability of the manuscript. After using this tool, the authors thoroughly reviewed and edited the content as needed and take full responsibility for the content of the published article.}

\bibliographystyle{IEEEtran}
\bibliography{references}

\section*{Acknowledgements}
The authors would like to acknowledge the support provided by the Deanship of Research (DR) at King Fahd University of Petroleum \& Minerals (KFUPM) for funding this work through project No. MbSC2601.

\section*{Author contributions}

Conceptualization: Umair Bin Waheed; 
Data curation: Ayrat Abdullin, Leo Eisner; 
Formal analysis: Umair Bin Waheed, Leo Eisner, Naveed Iqbal;
Methodology: Ayrat Abdullin, Umair Bin Waheed, Leo Eisner; 
Project administration: Ayrat Abdullin; 
Resources: Umair Bin Waheed, Leo Eisner; 
Supervision: Umair Bin Waheed; 
Validation: Leo Eisner, Naveed Iqbal;
Visualization: Ayrat Abdullin;
Writing - original draft: Ayrat Abdullin; 
Writing - review and editing: Ayrat Abdullin, Umair Bin Waheed, Leo Eisner, Naveed Iqbal.

\section*{Data availability}

The data supporting the findings of this study are available from Seismik s.r.o., but restrictions apply to their availability. These data were used under license for the current study and are therefore not publicly available. Data are, however, available from the authors upon reasonable request and with permission of Seismik s.r.o.

\section*{Competing interests}

The authors declare no competing interests.

\section*{Funding}

This research did not receive funding.

\section*{Additional information}

Correspondence and requests for materials should be addressed to U.B.W.

\begin{IEEEbiography}[{\includegraphics[width=1in,height=1.25in,clip,keepaspectratio]{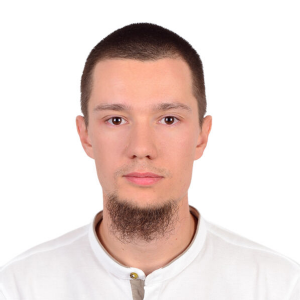}}]{Ayrat Abdullin} received the B.Sc. degree in geophysics from Kazan Federal University, Kazan, Russia, in 2010, and the M.Sc. degree in geosciences from Colorado State University, Fort Collins (CO), United States, in 2012. He then worked as a Research Assistant at the Crustal Deformation and InSAR Group (CDI), King Abdullah University of Science and Technology (KAUST), Thuwal, Saudi Arabia, and a Lead Specialist at the Technology Development Center, Tatneft Oil Company, Russia. 

Currently, Ayrat is a Ph.D. Candidate in Geophysics with Department of Geosciences at King Fahd University of Petroleum and Minerals (KFUPM), Dhahran, Saudi Arabia. His research interests include microseismic monitoring and machine learning applications in geophysics. 
\end{IEEEbiography}

\begin{IEEEbiography}[{\includegraphics[width=1in,height=1.25in,clip,keepaspectratio]{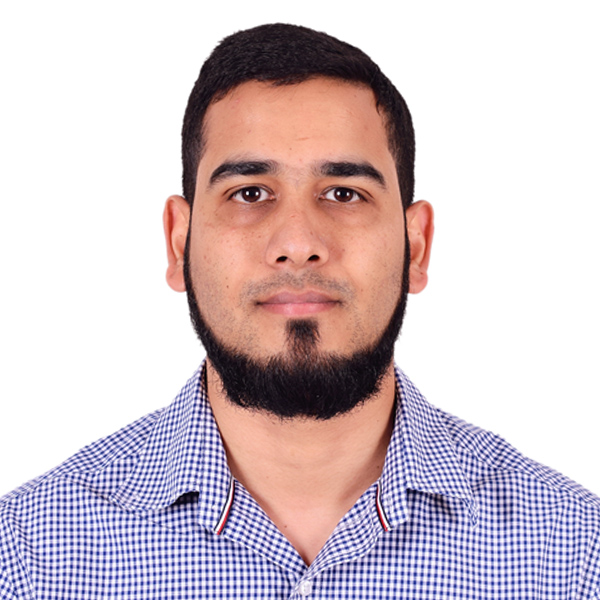}}]{Umair Bin Waheed} is an Associate Professor of Geophysics at King Fahd University of Petroleum and Minerals (KFUPM). His research interests are broadly in the area of computational geosciences. Umair graduated from KAUST with a Ph.D. under the supervision of Tariq Alkhalifah in 2015. Prior to joining KFUPM in 2017, he was a postdoctoral fellow at the Department of Geosciences, Princeton University and a Writing in Science and Engineering fellow at the Princeton Writing Program. Umair was the 2023 SEG Honorary Lecturer for Africa and Middle East on the topic of physics-informed machine learning. He also received the 2024 SEG Outstanding Educator Award.
\end{IEEEbiography}

\begin{IEEEbiography}[{\includegraphics[width=1in,height=1.25in,clip,keepaspectratio]{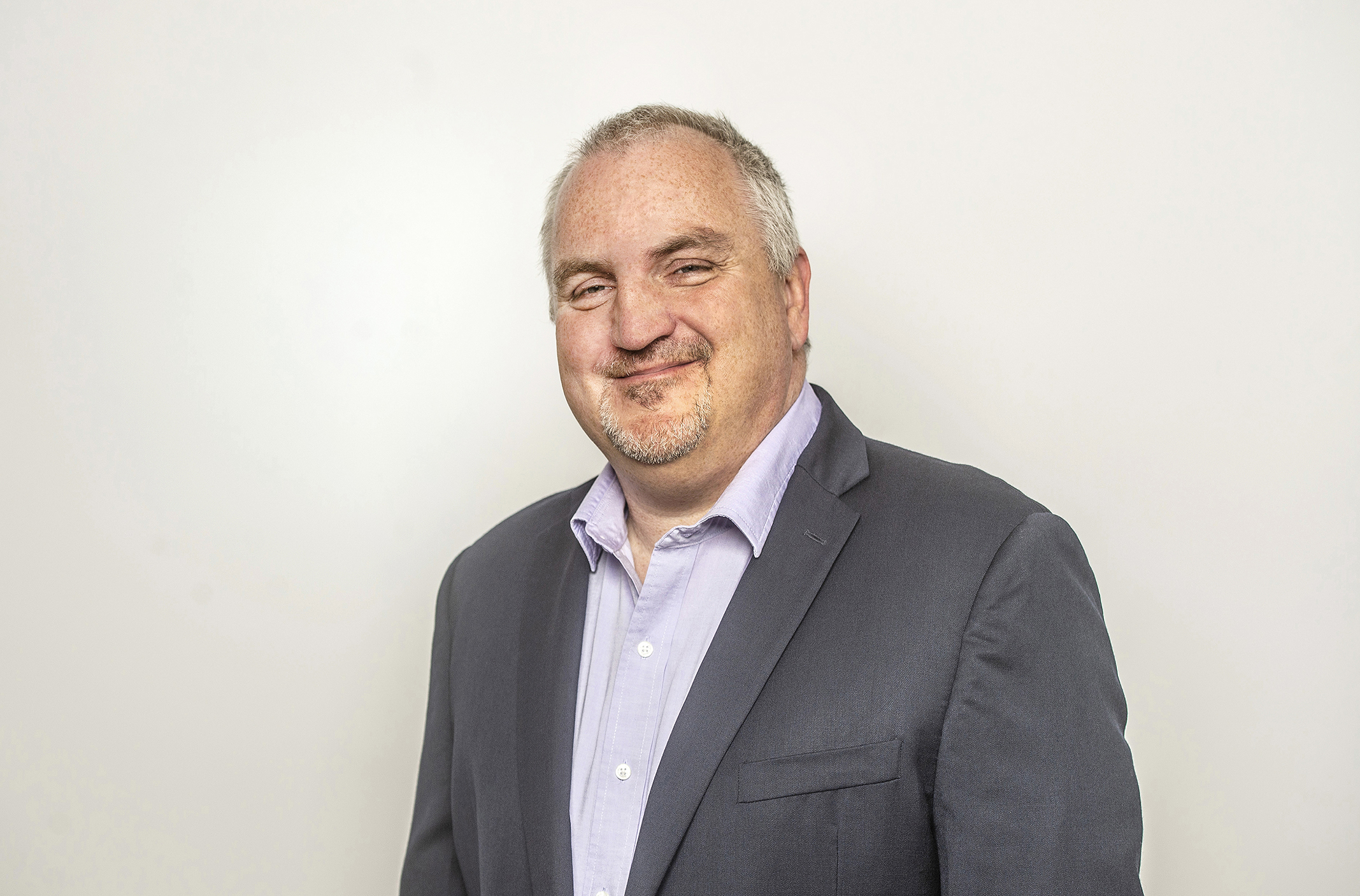}}]{Leo Eisner} After completing his Ph.D. on earthquake hazard assessment at the California Institute of Technology, Leo joined Cambridge Schlumberger Research as a Senior Research Scientist for six years. There he worked on down-hole monitoring of micro-seismic events. In 2008, Leo joined Microseismic Inc. (MSI), where he led the Development department, focusing on surface monitoring of microseismicity, in the capacity of Senior Geophysicist and Chief Geophysicist. Between 2010 and 2017 Leo carried out research as the Purkyne Fellow at the Institute of Rock Mechanics and Structure within the Academy of Sciences of the Czech Republic. Leo’s goal is to improve understanding of induced seismicity in general, as well as to help operators to understand and improve microseismic analysis in their reservoirs.
\end{IEEEbiography}

\begin{IEEEbiography}[{\includegraphics[width=1in,height=1.25in,clip,keepaspectratio]{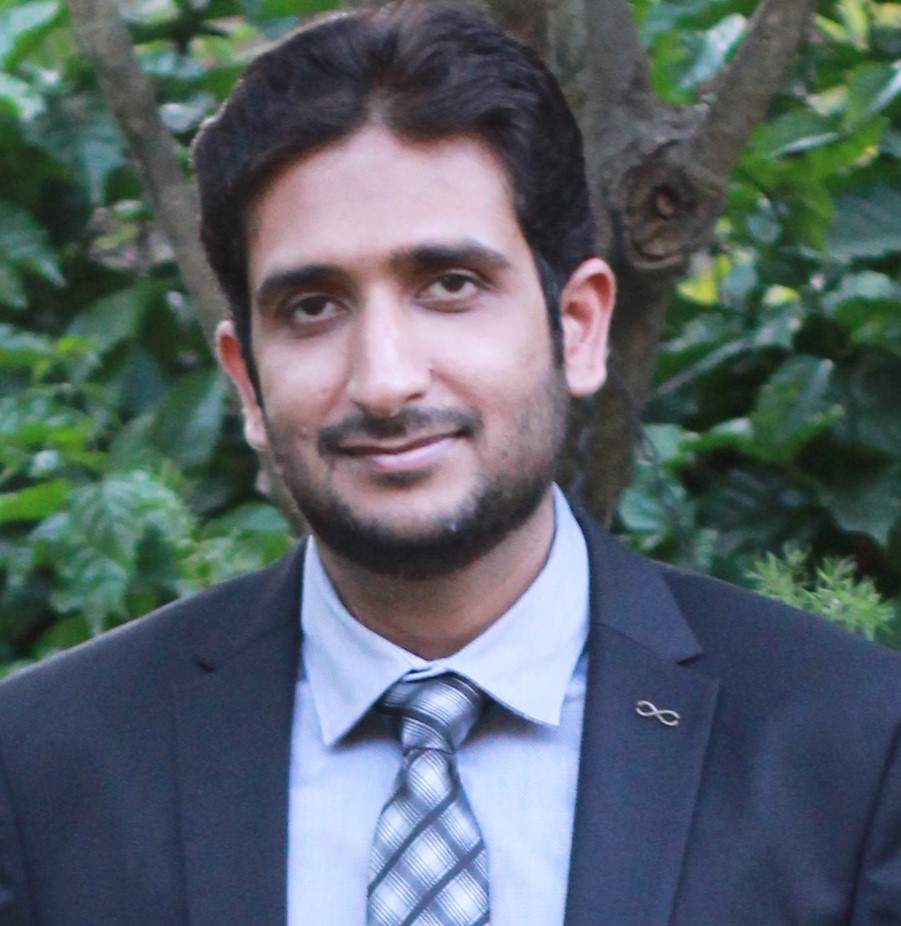}}]{Naveed Iqbal}, IEEE Senior Member, received the B.S. and M.S. degrees in electrical engineering from the University of Engineering and Technology, Peshawar, Pakistan, and the Ph.D. degree from the King Fahd University of Petroleum and Minerals (KFUPM), Saudi Arabia. He is currently an Assistant Professor at KFUPM. His research interests include adaptive algorithms, compressive sensing, heuristic algorithms, seismic signal processing, machine learning, and data acquisition networks.
\end{IEEEbiography}


\EOD

\end{document}